\documentclass[aps,pra,reprint,superscriptaddress,amsmath,amssymb,amsfonts]{revtex4-1}
\usepackage[pdftex]{graphicx}% Include figure files
\usepackage{dcolumn}% Align table columns on decimal point
\usepackage{bm}% bold math
\usepackage{epsfig}
\usepackage{pifont} 
\usepackage{multirow}
\usepackage{subfigure}
\usepackage{float}
\usepackage[utf8]{inputenc} 
\usepackage{epstopdf}
\usepackage{MnSymbol}

\newcommand{\ba}{\begin{eqnarray}}
\newcommand{\ea}{\end{eqnarray}}
\newcommand{\be}{\begin{equation}}
\newcommand{\ee}{\end{equation}}
\newcommand{\bea}{\begin{eqnarray}}
\newcommand{\eea}{\end{eqnarray}}

\newcommand{\T}{\mathrm{T}}
\newcommand{\Tr}{\mathrm{Tr}}

\newcommand{\Y}{\mathrm{Y}}

\newcommand{\abs}[1]{\ensuremath{\vert #1 \vert}}

\def\i {\mathfrak{i}}

\def\N {\mathbb{N}}

\usepackage{xcolor}

\usepackage{bbm}

\usepackage{amsthm}

\newtheorem{result}{Result}
\newtheorem*{main*}{Result}

\usepackage[bookmarks=false,pdfstartview={FitH}]{hyperref}

\allowdisplaybreaks

\begin{document}

\title{Fast computation of spherical phase-space functions of quantum many-body states}

\author{Bálint Koczor}
\email{balint.koczor@materials.ox.ac.uk}
\affiliation{University of Oxford, Department of Materials, Parks Road, Oxford OX1 3PH, United Kingdom}
\affiliation{Technische Universität München, Department Chemie, Lichtenbergstrasse 4, 85747 Garching, Germany}
\affiliation{Munich Center for Quantum Science and Technology (MCQST), Schellingstrasse~4, 80799 München, Germany}
%\email[Name]{Your e-mail address}
%\thanks{}
\author{Robert Zeier}
\email{r.zeier@fz-juelich.de}
\affiliation{Forschungszentrum Jülich GmbH, Peter Grünberg Institute, Quantum Control (PGI-8), 54245 Jülich, Germany}
\author{Steffen J. Glaser}
\email{glaser@tum.de}
\affiliation{Technische Universität München, Department Chemie, Lichtenbergstrasse 4, 85747 Garching, Germany}
\affiliation{Munich Center for Quantum Science and Technology (MCQST), Schellingstrasse~4, 80799 München, Germany}

\begin{abstract}
        Quantum devices are preparing increasingly more complex
        entangled quantum states. How can one effectively study 
        these states in light of their increasing dimensions?
        Phase spaces such as Wigner functions provide
        a suitable framework. We focus on 
        phase spaces for finite-dimensional
        quantum states of single qudits or 
        permutationally symmetric states of multiple qubits.
        We present methods to efficiently
        compute the corresponding phase-space
        functions which are at least an order 
        of magnitude faster than traditional methods.
        Quantum many-body states in much larger dimensions 
        can now be effectively 
        studied by experimentalist
        and theorists using these phase-space techniques.
\end{abstract}

\date{August 14, 2020}
%\date{\today}

%\keywords{
%%(PhySH terms)
%
%}

\maketitle

\section{Introduction}

Current (and near-term) quantum devices are expected to prepare increasingly more complex
entangled quantum states \cite{google_supr, catstate, Song574, preskill2018quantum}.
How can one effectively illustrate and analyze these states
in light of their increasing dimensions? Phase spaces \cite{SchleichBook,zachos2005,schroeck2013,Curtright-review}
such as Wigner functions have been widely used to meet this challenge.
We will focus in this work on representing
(finite-dimensional) quantum states of
single qudits or permutationally symmetric states of multiple qubits using spherical
phase spaces \cite{koczor2017,thesis}. 

Permutationally symmetric states include, e.g.,
Greenberger–Horne–Zeilinger (GHZ) and squeezed states, and they
have immediate applications in quantum metrology for optimally estimating, e.g.,
magnetic field strengths \cite{review,toth14,giovannetti11,koczor2019variational}.
Phase spaces are a useful tool for
visualizing experimentally generated quantum many-body states of
atomic ensembles \cite{google_supr, catstate,mcconnell2015,haas2014},
Bose-Einstein condensates \cite{anderson1995,ho1998,ohmi1998,stenger1999,lin2011,treutlein2010,Schmied2011,hamley2012,strobel2014},
trapped ions \cite{leibfried2005,bohnet2016,monz2011}, and light polarization \cite{bouchard2016,klimov2017,chaturvedi2006}.
On the theoretical side, phase spaces  provide the necessary intuition
as they naturally reduce to classical phase spaces 
in the limit of a vanishing Planck constant \cite{Gro46,Moy49,1bayen1978,2bayen1978,berezin74,berezin75}.
Such phase-space techniques, and related quantization methods  \cite{Wey27,Weyl31,Weyl50}, also play a vital role
in harmonic analysis and in the theory of pseudo-differential operators
\cite{thewignertransform,bornjordan,groechenig2001foundations,cohen1966generalized,Cohen95}.

%-------------------------
\addtocounter{footnote}{1}
\footnotetext[\value{footnote}]{
	We computed Wigner functions of tensor operators of high rank $j>1$,
	whose functional form we also know analytically as spherical harmonics
	-- these decompose into a large number of non-trivial Fourier components.
	}
\newcounter{footfunc}
\setcounter{footfunc}{\value{footnote}}
%-------------------------
\begin{figure*}[tb]
\begin{centering}
\includegraphics{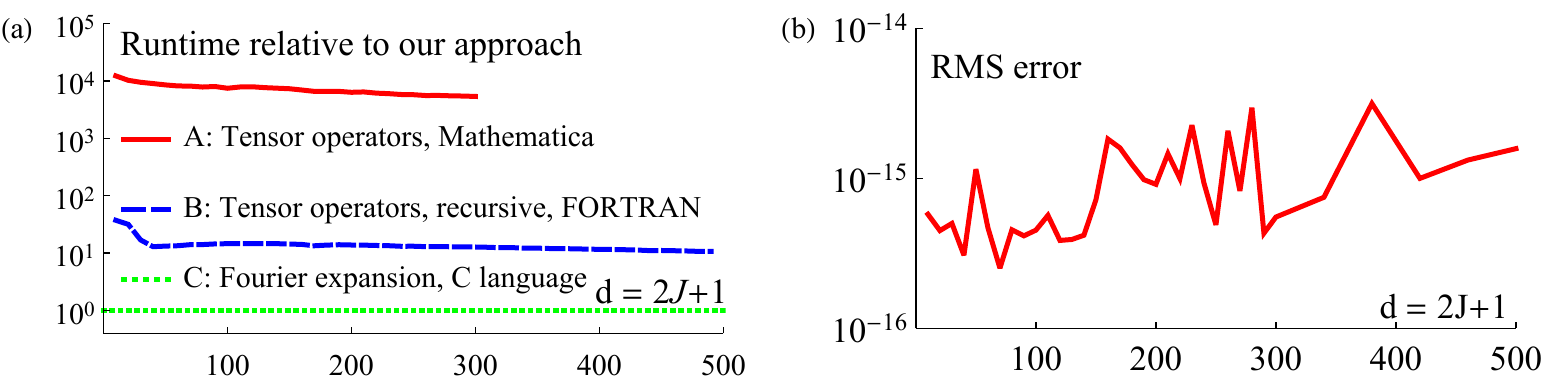}
\caption{%
(a) Run times of earlier Methods A and B relative
to our Method C for computing phase-space functions of
quantum states with an increasing dimension $d  = 2J{+}1$
are at least an order of magnitude slower.
Methods A and B both employ tensor-operator decompositions (Sec.~\ref{prior}).
Method A relies on Mathematica's built-in method to compute Clebsch-Gordan coefficients and
Method B uses an efficient, recursive algorithm \cite{schulten1,schulten2,wsymbols,gitwigsymb}. 
Our Method C (Sec.~\ref{mainsec}) combines 
spherical sampling techniques \cite{driscoll94,kennedy2013book},
explicit descriptions of rotation operators \cite{tajima2015analytical,wigdmatrix2},
Fourier series expansions, 
and fast Fourier transforms (FFT).
The run times depend only on $d$ and not the quantum state.
(b) Root mean square (RMS) errors for certain quantum states \cite{Note\thefootfunc}
relative to their analytically known formula.
Method C shows a high numerical precision comparable to machine precision.
\label{precisionfig}
}
\end{centering}
\end{figure*}
%-------------------------

In this work, we consider spherical phase spaces of finite-dimensional quantum states
and we develop a novel approach to efficiently compute these phase-space representations.
For up to which dimensions can phase spaces be practically utilized?
Our approach has a significant advantage in this regard as
it allows for much larger dimensions to be addressed in a reasonable time frame.
Therefore, phase-space descriptions of quantum many-body states
are now feasible for dimensions which were beyond the reach
of prior approaches. In summary, our results will enable practitioners and experimentalist---but 
also theorists---to visualise and study complex quantum states in 
considerably larger dimensions.

This is accomplished by applying 
an efficiently computable Fourier series expansion
and a fast Fourier transform (FFT) \cite{AM04}. 
In particular, 
Fig.~\ref{precisionfig}(a) compares the 
the run time of our Method C
(as detailed in Sec.~\ref{mainsec}) to the traditional Methods A and B (see Sec.~\ref{prior})
and, indeed, our Method C is at least an order of magnitude faster.
Moreover, Fig.~\ref{precisionfig}(b) highlights that the root-mean-square error
of certain test cases
is comparable to machine
precision for the considered dimensions and this suggests 
that our approach is numerically stable.
We provide implementations in various programming environments (see Sec.~\ref{sourceCode} and \cite{gitcode}),
including C~\cite{kernighan2006c}, MATLAB~\cite{MATLAB:2019}, Mathematica~\cite{Mathematica},
and Python~\cite{van1995python}.

Our work has the following structure: 
We first discuss our motivation 
and highlight applications in Sec.~\ref{applications}.
Prior computational 
approaches to determine phase-space representations
of finite-dimensional quantum systems are considered in 
Sec.~\ref{prior}. In order to set the stage,
we shortly recall the parity-operator description of spherical phase spaces
which we have developed in \cite{koczor2017}.
Section~\ref{mainsec} constitutes the main part of our manuscript
where we develop our novel approach to efficiently compute
spherical phase-space representations up to arbitrarily fine resolutions.
We continue with a discussion of our results and further applications
in Sec.~\ref{discussion}, before we conclude. Important details 
are explained in appendices.

\section{Motivation and Applications\label{applications}}

Various quantum-technology efforts (such as quantum computing or metrology)
aim at creating large
entangled multi-qubit states. Here, we focus in particular on the important
class of states that are symmetric under permutations of qubits.
These states include important families such as GHZ or squeezed states
which are central in, e.g., quantum metrology \cite{review}
or entanglement verification \cite{catstate,Song574}.  They  are also
typically illustrated and analyzed in their phase-space representation  (see, e.g.,
\cite{Song574, review}) which can be naturally plotted on the surface of a sphere.
This is reflected by the inherent symmetries and reduced degrees of freedom
as compared to general multi-qubit states. Before starting the technical
discussion in Sec.~\ref{prior}, we will now motivate our topic and 
highlight applications.

We first recall that permutationally symmetric states
with $N = 2J$ qubits can be mapped to states of 
a single spin $J$ (or qudit with $d=2J{+}1$) where $J$ denotes
a positiv integer or half-integer
 \cite{KZG,thesis,Dicke1954,stockton2003,toth2010,lucke2014}.
Permutation symmetry appears in various applications including
probe states in quantum metrology 
for optimal sensing, e.g., magnetic fields \cite{review,toth14,giovannetti11,koczor2019variational}.
Permutationally symmetric qubit states can be efficiently reconstructed and are used for entanglement
verification \cite{toth2010,stockton2003,lucke2014, review, catstate,Song574}.
We will illustrate a few practically relevant, high-dimensional examples
for which traditional methods (see Sec.~\ref{prior})
take an impractically large amount of time in order to determine the desired
phase-space function. Further discussions and 
applications are deferred to
Sec.~\ref{discussion}.

%-------------------------
\begin{figure*}[tb]
\begin{centering}
\includegraphics[width=\textwidth]{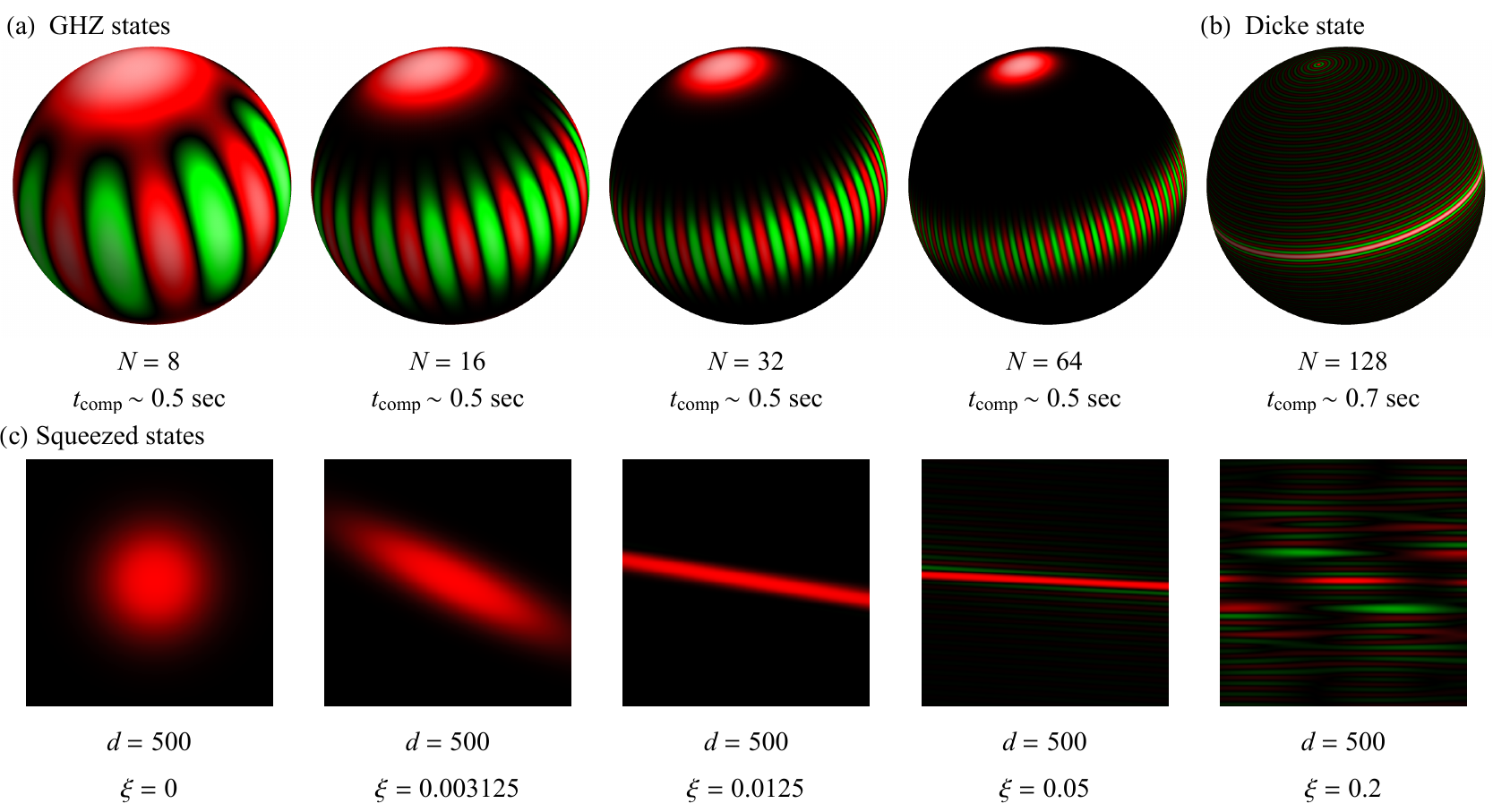}
\caption{%
Applications highlighted by numerically computed Wigner functions 
of single-qudit states with $d=2J{+}1$ which are equivalent to permutation-symmetric states of $N=d{-}1$ qubits:
(a) GHZ states for $N \in \{8, 16, 32, 64\}$, (b) Dicke state $|J m\rangle$ for $N=128$ and $m=0$.
The run time (without 3D graphics and rasterization) using Method D (Sec.~\ref{mainsec}) 
on a laptop in Mathematica is dominated by the FFT on a $1024 \times 1024$ grid.
(c) Similarly, Wigner functions of squeezed states $|\xi \rangle := \exp[- i \xi \, \mathcal{I}_x^2] |0\rangle^{\otimes N}$
with varying squeezing angle $\xi$ and fixed $d=N{+}1=500$ in
a plane for a small spherical subset with $\abs{\theta} \leq 0.05 \pi$
(run time $\approx 1$ min):
Gaussian for $\xi=0$ (left);
squeezed Gaussians for $\xi < 0.05$. Larger $\xi \geq 0.05$
lead to non-trivial and rapidly oscillating shapes which are nicely recovered,
while analytical approximations fail in this regime.
Red (dark gray) and green (light gray) for positive and negative values, respectively. The brightness
indicates the absolute value of the function relative to its global maximum.
\label{example}
}
\end{centering}
\end{figure*}
%-------------------------

The first example considers and highlights the  Greenberger–Horne–Zeilinger (GHZ) state 
$(|0\rangle^{\otimes N} +|1\rangle^{\otimes N})/\sqrt{2}$
as the superposition of the all-zero and all-one state for $N$ qubits
which can be interpreted as the spin-up and spin-down states of a single qudit.
Their high degree of entanglement 
supports
the ultimate quantum precision in metrology, which is known as the Heisenberg limit \cite{review}.
GHZ states have been successfully created in numerous experiments with, e.g.,
trapped ions \cite{monz2011}, superconducting qubits
\cite{Song574}, and Rydberg atoms \cite{catstate} for up to 20 qubits.
Although phase-space functions 
of GHZ states can be analytically approximated
for large dimensions \cite{koczor2017,KZG}, we are interested in 
computing them exactly 
within numerical precision and without relying on approximations.
Figure~\ref{example}(a) shows Wigner functions of 
GHZ states for an increasing 
number of qubits with $N \in \{8, 16, 32, 64\}$. Already the case $N=32$ is currently beyond
the experimental state of the art \cite{catstate,Song574},
but near-term quantum hardware are expected to deliver GHZ states of larger dimensions
via, e.g., linear-depth quantum circuits \cite{preskill2018quantum}.

We also consider so-called symmetric Dicke states \cite{Dicke1954} which are defined
\cite{thesis,stockton2003} as a superposition of all permutations
of computational basis states with a fixed number 
of zeros and ones in a multi-qubit system. In particular,
\begin{equation} \label{perminvariantstates}
|Nn\rangle :=  \tfrac{1}{\sqrt{p}} \, \sum_{k=1}^{p} P_k  \, | \underbrace{1, 1, \dots, 1}_{n},\underbrace{0, \dots, 0}_{N-n} \rangle ,
\end{equation}
where the sum runs over all $p=\binom{N}{n}$ distinct permutations $P_k$ of the $N$ qubits.
These states are isomorphic to the single-qudit states $|J m\rangle$ by mapping $N$ to  $J/2$
and $m$ to $(N/2-n)$. We plot the Dicke state $|J m\rangle$ with $d=2J{+}1=129$ and
$m=0$ in Fig.~\ref{example}(b). This corresponds to a highly entangled quantum state of 128 indistinguishable
qubits where 64 qubits are in the $|0\rangle$ state and 64 qubits are in the $|1\rangle$ state
(refer to Eq.~\eqref{perminvariantstates}). One observes an axial symmetry (i.e. invariance under 
global $Z$ rotations) and strong entanglement
results in heavily oscillating Wigner functions in  Fig.~\ref{example}(b).

Finally, squeezed states $|\xi \rangle := \exp[- i \xi \, \mathcal{I}_x^2] |0\rangle^{\otimes N}$
are obtained from the spin-up state of a single qudit or, equivalently, the all-zero state of $N$ qubits
under the influence of a squeezing interaction Hamiltonian $\mathcal{I}_x^2$. The corresponding evolution time
$\xi$  is known as the squeezing angle \cite{ma2011quantum} and
$\mathcal{I}_x$ is the $x$ component of the 
total angular momentum operator, i.e., 
proportional to the sum of all Pauli $\sigma_x$ operators that act on different qubits.
These states have been created in various experiments including Bose-Einstein condensates
\cite{anderson1995,ho1998,ohmi1998,stenger1999,lin2011,treutlein2010,Schmied2011,hamley2012,strobel2014,hosten2016measurement}
for up to thousands of atoms.
In such experiments,
these finite-dimensional squeezed states
correspond to internal
degrees of freedom (which we treat as an effective qudit) of fundamentally indistinguishable atoms.
We plot their Wigner functions for the case of $d=N+1=500$
and an increasing squeezing angle $\xi$ in Fig.~\ref{example}(c).
For such large dimensions squeezed states with small squeezing angles 
can be approximated well using the techniques
described in \cite{koczor2017, KZG}. In particular,
the spin-up state $|0\rangle^{\otimes N}$ for $\xi=0$  in Fig.~\ref{example}(c) is a Gaussian-like function
because the sphere can be approximated locally as a plane. For small squeezing angles,
these states can be analytically approximated using star products \cite{KZG,klimov2005classical}.
Their phase-space representations are squeezed Gaussian functions which are very similar to the ones known in quantum 
optics \cite{ma2011quantum,Leonhardt97}. This is illustrated in Fig.~\ref{example}(c) where
the aforementioned approximations apply to the cases $\xi=0$, $\xi=0.003125$, and $\xi= 0.0125$. 
For larger squeezing angles, Wigner functions 
will, however, deviate strongly from simple
squeezed Gaussian states and non-trivial, heavily oscillating contributions become dominant
as is shown in Fig.~\ref{example}(c) for $\xi=0.05$ and $\xi= 0.2$.
This motivates our numerical approach to \emph{exactly} determine phase-space functions for large spin-like systems
(and permutationally symmetric multi-qubit states) where analytical approximations do usually fail.

\section{Traditional methods to compute spherical phase-space functions\label{prior}}

We now discuss traditional methods 
to compute phase-space functions of
qudit states with $d=2J{+}1$
and consider
the full class of 
$s$-parametrized phase spaces with $-1 \leq s \leq 1$.
This includes Wigner functions ($s=0$) \cite{dowlingagarwalschleich}, 
Husimi $Q$ functions ($s=-1$) \cite{agarwal81}, and Glauber $P$ functions ($s=1$).
Spherical phase spaces are parametrized
by two Euler angles $(\theta,\phi)$
with $0 \leq \theta \leq \pi$ and $0 \leq \phi < 2\pi$.
Building on the pioneering work by Agarwal \cite{agarwal81,dowlingagarwalschleich}, 
$s$-parametrized
phase-space functions \cite{koczor2017}
\begin{equation}
\label{sphdecomp}
F_\rho (\theta,\phi,s) 
= \tfrac{1}{R} \, \sum_{j=0}^{2J} \sum_{m=-j}^{j}   (\gamma_j)^{-s} \, c_{jm} \, \Y_{jm}(\theta,\phi)
\end{equation}
can be expanded 
into spherical harmonics $\Y_{jm}(\theta,\phi)$ \cite{Jac99}.
The constant $\gamma_j:=R\, \sqrt{4\pi} (2J)! \, [ (2J{+}j{+}1)! \, (2J{-}j)! \,  ]^{{-}1/2}$ 
and the spherical radius $R:=\sqrt{J/(2\pi)}$ are used in Eq.~\eqref{sphdecomp}.
The expansion coefficients $c_{jm} := \Tr \,[ \rho\,  \T_{jm}^\dagger ]$
are computed from the density matrix $\rho$
and the tensor-operator coefficients $\T_{jm}$ 
\cite{racah42,fano59,silver76,CH98}.  The matrix elements
\begin{subequations}
\label{topelements}
\begin{align}
[\T_{jm}]_{m_1 m_2} & = 
\sqrt{(2j{+}1)/(2J{+}1)} \, C^{J m_1}_{J m_2, j m}\\
& =(-1)^{J-m_2}\, C^{jm}_{Jm_1,J,-m_2}
\end{align}
\end{subequations}
are determined by 
Clebsch-Gordan coefficients $C^{J m_1}_{J m_2, j m}$
where $m_1,m_2 \in \{J,\ldots,-J\}$ \cite{messiah1962b,Brif98,brif97,BL81,Fano53}.

Equation~\eqref{sphdecomp} describes 
the standard approach for numerically computing spherical phase-space functions.
In a first step, it relies on efficient approaches to calculate Clebsch-Gordan coefficients. 
The calculation of the expansion coefficients $c_{jm}$
is, however, computationally expensive for large dimensions  $d=2J{+}1 \gg 1$.
In particular, one needs to determine $\mathcal{O}(d^2)$ distinct tensor-operators $\T_{jm}$
and their matrix entries.
Appendix~\ref{topdecomp} clarifies that  $\mathcal{O}(d^3)$ Clebsch-Gordan
coefficients have to be calculated which dominates the run time for computing all of the 
$\mathcal{O}(d^2)$ expansion coefficients $c_{jm}$ in Eq.~\eqref{sphdecomp}.

\begin{figure*}[tb]
	\begin{centering}
		\includegraphics{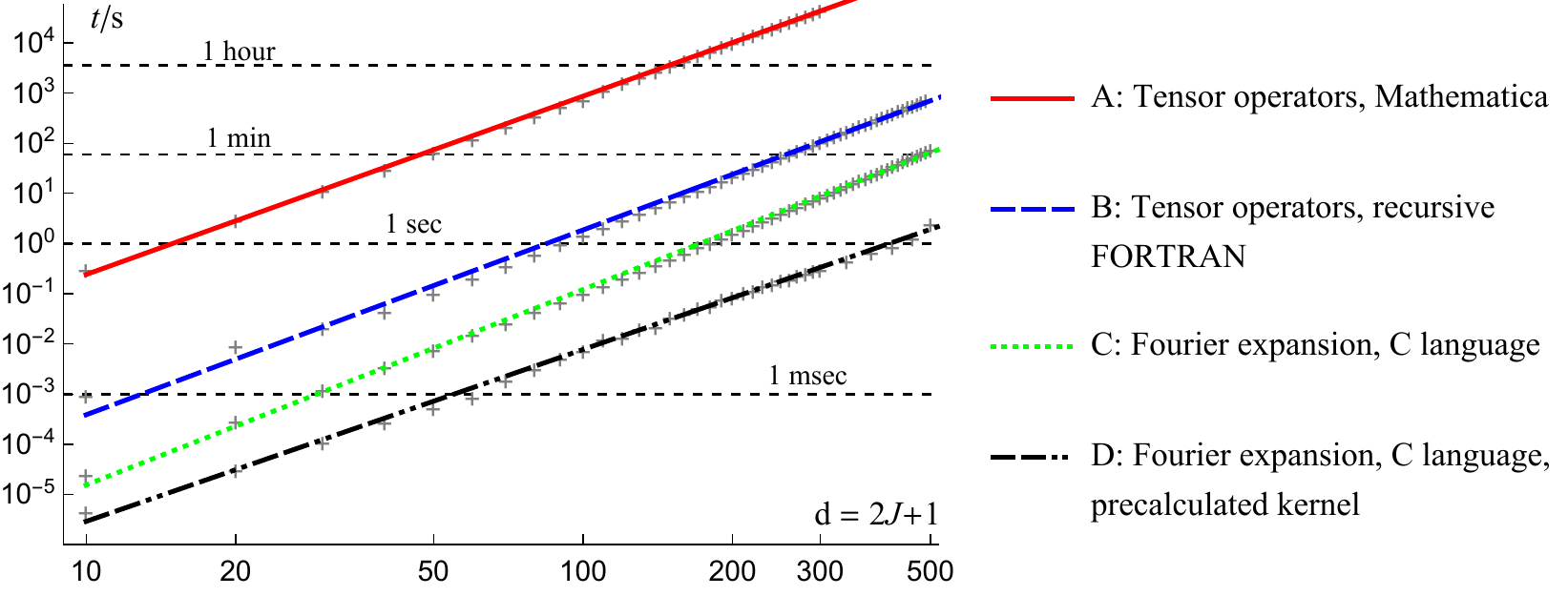}
		\caption{%
		        Run times for computing phase-space functions 
		        with Methods A to D (Sec.~\ref{prior} and \ref{mainsec}) for 
		        dimensions $d = 2J{+}1 \leq 500$, while ignoring asymptotically negligible
		        contributions from spherical harmonics transformations (for A and B) or FFTs (for C and D).
		        Methods A, B and C show a similar asymptotic behavior $\mathcal{O}(d^4)$.
		        But our Method C is at least an order of magnitude faster (using C code)
		        which allows for much larger dimensions. Building on Method C,
		        Method D is even faster and has a lower asymptotic time complexity
		        $\mathcal{O}(d^3)$ but relies on precomputations 
		        and additional disk storage (Table~\ref{storage}). 
		        The run times depend only on $d$ and not the quantum state
		        \cite{Note\thefootxeon}.
			\label{timesfigure}
		}
	\end{centering}
\end{figure*}

Two different approaches to calculate Clebsch-Gordan coefficients 
result in two different methods (Method A and B) to the coefficients $c_{jm}$.
Method A uses the built-in Mathematica~\cite{Mathematica}
function that performs arbitrary-precision integer arithmetic.
In Method B, the run time can be significantly reduced by numerically
computing Clebsch-Gordan coefficients using a FORTRAN~\cite{backus1964fortran} 
implementation~\cite{gitwigsymb} of a recursive algorithm~\cite{schulten1,schulten2,wsymbols}.
Methods A and B are compared in Fig.~\eqref{timesfigure}. For Method A (B), 
all tensor operators for certain dimensions $d\leq 300$ ($d\leq 500$) have been determined
and we estimate a complexity $\mathcal{O}(d^4)$ in this range.

After the expansion coefficients $c_{jm}$ have been obtained, 
the phase-space function $F_\rho (\theta,\phi,s)$ is 
spherically sampled in a second step 
by applying a fast spherical harmonics transform
which might rely on equiangular samples or Gauss-Legendre grids.
The second step requires a practically and asymptotically 
negligible time of $\mathcal{O}(d^3)$ when compared to the first step.
Spherical harmonics transforms are widely used in various  scientific contexts and efficient
implementations are available
\cite{suda2002fast,driscoll94,libsharp,shtns,mohlenkamp1999fast}.

%-------------------------
\addtocounter{footnote}{1}
\footnotetext[\value{footnote}]{
All data points were obtained on a desktop computer with an Intel\textsuperscript{\tiny\textregistered}
Xeon\textsuperscript{\tiny\textregistered} W-2133 processor at 3.60GHz 
using a single thread.}
\newcounter{footxeon}
\setcounter{footxeon}{\value{footnote}}
%-------------------------

\section{The parity-operator description of spherical phase spaces\label{parity}}

We recall the parity-operator description of spherical phase spaces
developed in \cite{koczor2017} in order to develop faster methods 
to compute spherical phase-space functions in Sec.~\ref{mainsec}.
We keep the notation introduced in Sec.~\ref{prior}
and specify the rotation operator as
$\mathcal{R}(\theta,\phi):= e^{i\phi \mathcal{J}_z} e^{i\theta \mathcal{J}_y} $,
where
$\mathcal{J}_z$ and $\mathcal{J}_y$ are components of the 
angular momentum operator \cite{messiah1962a}.
Building on \cite{stratonovich,agarwal81,vgb89,Brif98,brif97}, 
the $s$-parametrized
phase-space functions are defined in \cite{koczor2017}
as expectation values of rotated parity operators $M_s$ by
\begin{equation}
\label{PSrepDefinition}
F_\rho (\theta,\phi,s) :=  \Tr \,[ \, \rho \, \mathcal{R}(\theta,\phi)  M_s  \mathcal{R}^{\text{\emph{$\dagger$}}}(\theta,\phi) ].
\end{equation}
This extends work \cite{heiss2000discrete,KdG10,tilma2016,rundle2017,RTD17} 
on rotated parity operators to all $s$-parametrized phase spaces.
The parity operator
\begin{equation}
\label{DefOfspinParityoperators}
M_s := \tfrac{1}{R} \, \sum_{j=0}^{2J} \sqrt{\tfrac{2j{+}1}{4 \pi}} (\gamma_j)^{-s} \, \T_{j0},
\end{equation}
is defined by its expansion into diagonal tensor operators $\T_{j0}$ 
of order zero. The corresponding matrix elements
are given by $[\T_{j0}]_{mm'}= 
\delta_{mm'} \sqrt{ (2j{+}1) /(2J{+}1) } \, C_{Jm,j0}^{Jm}$ for
$j\in \N {\cup} \{0\}$
and $m,m'\in\{-J,\ldots,J\}$. 
Equation~\eqref{sphdecomp} could be recovered by applying the
rotation operators to the tensor operators in
Eq.~\eqref{DefOfspinParityoperators} as
$\mathcal{R}(\theta,\phi)  M_s  \mathcal{R}^{\text{\emph{$\dagger$}}}(\theta,\phi)
= 
\tfrac{1}{R} \, \sum_{j=0}^{2J} \sum_{m=-j}^{j}  (\gamma_j)^{-s} \, \T^\dagger_{jm} \Y_{jm}(\theta,\phi)$.

For an increasing spin number $J$, spherical phase spaces converge to their
infinite-dimensional counterparts while rotations transform into translations along
the tangent of a sphere \cite{thesis,koczor2017,koczor2018,KZG}.
While we focus here on single qudits (and permutationally symmetric quantum states of multiple qubits), 
generalizations of the parity-operator approach to arbitrary coupled quantum states
are also available \cite{gzg15,leiner2020symmetry,tilma2016,koczor2016,rundle2017}.

\section{Efficient computation of spherical phase-space functions \label{mainsec}}

We develop now our main results on efficiently computing spherical phase-space 
functions. Section~\ref{directscheme} presents a first approach
using parity operators (see Sec.~\ref{parity}), an explicit form for rotation operators, and a spherical sampling 
strategy. This does---by itself---not lead to an effective approach. But it provides
the necessary ingredients to specify spherical phase-space functions
as a finite Fourier series in Sec.~\ref{main} which includes our efficient algorithm for 
the corresponding Fourier coefficients.
A fast Fourier transform is then applied as
detailed in Sec.~\ref{fftsec} to recover 
an equiangular  spherical sampling 
of the phase-space function.
Finally, we discuss  implementations
of our efficient algorithms in Sec.~\ref{sourceCode}.

\subsection{A first approach via parity operators, matrix entries of rotations,
 and spherical sampling\label{directscheme}}

Equation~\eqref{PSrepDefinition} can be directly applied to 
calculate phase-space functions as expectation values of
rotated parity operators. The parity operators 
are determined by Eq.~\eqref{DefOfspinParityoperators}
and the matrix entries of the rotation operator 
$[\mathcal{R}(\theta,\phi)]_{m_1m_2} = D^J_{m_1 m_2} (\theta,\phi)$ \cite{BL81}
are analytically given as Wigner-D functions
(which are widely available in software environments such as Mathematica).
We also use results of \cite{tajima2015analytical,wigdmatrix2}
to compute the matrix entries of the rotation operator
using fast Fourier transforms (see Appendix~\ref{appendixRotationFourier}).
The phase-space function is then computed as the trace of the matrix product
of the operators in \eqref{PSrepDefinition}. 

One additional part in this first approach is 
the equiangular spherical sampling scheme of \cite{driscoll94,kennedy2013book}.
As phase-space functions
are band limited ($0 \leq j\leq2J$) with regard to 
their spherical harmonics decompositions,
we can apply spherical sampling schemes with
a discretized grid of spherical angles $(\theta_k,\phi_\ell)$.
One can uniquely represent a phase-space function by sampling 
on an equiangular grid 
\begin{equation} \label{sphsamples}
(\theta_k={\pi k}/{n},\phi_\ell={2\pi \ell}/{n}) \;\text{ for }\; k, \ell \in \{ 0,\dots,n{-}1\}
\end{equation}
with $n^2\geq (4J{+}2)^2=(2d)^2$ rotation angles \cite{driscoll94,kennedy2013book}.
One then evaluates Eq.~\eqref{PSrepDefinition}
at all angles in Eq.~\eqref{sphsamples} to obtain a equiangular spherical
sampling of the phase-space function. 
However,  this first approach requires
matrix multiplications for each of the $\mathcal{O}(d^2)$ spherical angles.
This leads to inefficiencies and
an overall run time of $\mathcal{O}(d^m)$, where $4.2 \lessapprox m \leq  5$
depending on the efficiency of the
matrix-multiplication algorithm (and $m=5$ corresponds to a naive implementation)
\footnote{We remark that when implementing this approach, one should choose a minimal resolution of $N = 2d$.
	After performing the computation, one can refine the resolution by Fourier transforming the result,
	then zero filling it, and finally applying an inverse Fourier transform.
}.
More effective methods are presented in  Sec.~\ref{main}.
The presented approach can be combined
with the algorithm of
\cite{driscoll94,kennedy2013book} to recover the spherical-harmonics expansion coefficients $c_{jm}$ in 
Eq.~\eqref{sphdecomp}.

\subsection{Efficient algorithms for the Fourier coefficients \label{main}}

We now expand on the approach in Sec.~\ref{directscheme} 
by exploiting the structure of the
rotated parity operators and by analytically evaluating the matrix products in Eq.~\eqref{PSrepDefinition}.
This facilitates a novel computational scheme for computing the Fourier expansion
of spherical phase-space functions which significantly differs from the methods
in \cite{tajima2015analytical,wigdmatrix2}. We begin by computing 
the Fourier expansion coefficients of the rotation operators $\mathcal{R}(\theta,\phi)$.
Recall that any (unitary) matrix can be written in terms of its spectral resolution
which also holds for
\begin{equation} \label{rotopresolEQ}
\mathcal{R}(\theta,\phi)= e^{i\phi \mathcal{J}_z} e^{i\theta \mathcal{J}_y} 
=
\sum_{\ell,m=-J}^{J} e^{i \ell \theta} e^{i  m \phi}   \, A_\ell  B_m.
\end{equation}
As detailed in Appendix~\ref{appendixRotationFourier}, 
$A_\ell$ and  $B_m$  are projection operators that project
onto the eigenvectors of the spin operators $\mathcal{J}_y$ and $\mathcal{J}_z$,
respectively. 
The dependence on the rotation angles has been completely absorbed
into the Fourier components $e^{i \ell \theta} e^{i  m \phi}$.

We can now analytically evaluate the trace of matrix products in Eq.~\eqref{PSrepDefinition}
and we prove in Appendix~\ref{derivation} that the phase-space function 
\begin{equation} \label{PSfctFourier}
F_\rho (\theta,\phi,s) = \sum_{\ell,m= -2J}^{2J} 
e^{i \ell \theta}  e^{i m \phi} F_{\ell m}
\end{equation}
can
be decomposed into a finite, band-limited Fourier series.
The Fourier expansion coefficients $F_{\ell m}$ 
implicitly depend on the
density matrix $\rho$ and the parity operator $M_s$ (as well as $s$) and
they can be obtained from $\rho$ via
a linear transformation:
\begin{result} \label{result1}
	The Fourier expansion coefficients in Eq.~\eqref{PSfctFourier} 
	of a spherical phase-space function $F_\rho (\theta,\phi,s)$ of
	a quantum state $\rho$ of dimension $d = 2J{+}1$
	are given by
	\begin{equation} \label{fouriercoeffsmain}
	F_{\ell m} = \sum_{ \lambda = \max(-J, -J{-}m)}^{\min(J, J{-}m)}
	\rho_{\lambda,\lambda+m}  \, [K_\ell]_{\lambda,\lambda+m},
	\end{equation}
	where $-2J \leq \ell,m \leq 2J$ and $\rho_{m_1,m_2}:= \langle Jm_1 |\rho | J m_2\rangle$ 
	are the density-matrix entries in
	the standard qudit basis. 
\end{result}

A proof of Result~\ref{result1} is given in Appendix~\ref{derivation}.
The transformation matrices $K_\ell \in \mathbb{C}^{d\times d}$ implicitly depend on 
the parity operator $M_s$ (and $s$).
They can be efficiently calculated as a finite sum (see Appendix~\ref{precalc})
\begin{equation} \label{precaleq}
K_\ell = \sum_{ \nu = \max(-J, -J{-}\ell)}^{\min(J, J{-}\ell)}  [\tilde{M}_s ]_{\nu,\nu + \ell}  \, | U_\nu \rangle \langle  U_{\nu + \ell}|.
\end{equation}
Here, $\tilde{M}_s$ denotes the  parity operator $M_s$ transformed into the
eigenbasis of the operator $\mathcal{J}_y$,  and $| U_\nu \rangle$ are the eigenvectors
of $\mathcal{J}_y$, such that $\mathcal{J}_y | U_\nu \rangle := \nu | U_\nu \rangle$.
The matrix entries of $\tilde{M}_s$ are therefore given as 
$[\tilde{M}_s ]_{a b} =   \langle U_a | M_s | U_b \rangle $.

Result~\ref{result1} leads to two different algorithms to
compute the Fourier coefficients in Eq.~\eqref{PSfctFourier}
(as detailed in Appendix~\ref{precalc}).
These algorithms are then combined with a fast Fourier transform
(which has a much smaller run time)
 in order to effectively compute  
an equiangular  spherical sampling 
of the spherical phase-space function (as discussed in Sec.~\ref{fftsec}).
The first algorithm to compute the Fourier coefficients
is denoted as Method~C: The
transformation matrix $K_\ell$ is computed for a fixed $\ell$ via Eq.~\eqref{precaleq}
in $\mathcal{O}(d^3)$ time. Then, $K_\ell$ 
is used to compute  the Fourier coefficients $F_{\ell m}$
for a fixed $\ell$ via \eqref{fouriercoeffsmain} in
$\mathcal{O}(d^2)$ time (which is less than
the previous step). This is repeated for every $\ell \in \{-J, \dots, J\}$.
Computing $F_{\ell m} $ takes overall
 $\mathcal{O}(d^4)$  time and $\mathcal{O}(d^2)$ memory.
 
The run time of a C implementation of Method~C is compared in Fig.~\ref{timesfigure}
to the traditional Methods~A and B from Sec.~\ref{prior}.
We empirically observe an asymptotic scaling of $\mathcal{O}(d^4)$
for all three methods and $d \leq 500$, which
is visible as near-parallel lines in the log-log plot of Fig.~\ref{timesfigure}.
However, Method~C is evidently much faster. 
Figure~\ref{precisionfig} (a) shows the 
relative runtimes of Methods~A and B compared to Method~C
highlighting that Method~C
is at least an order of magnitude faster. Consequently,
Method~C can be used for much larger dimensions.

\begin{table} 
\caption{Disk storage, RAM, and computing times for Methods C and D
(Result~\ref{result1} and Fig.~\ref{timesfigure}) 
with empirical complexities $\mathcal{O}(d^k)$;
matrices $K_\ell$ are computed on-the-fly (C) or have been precomputed (D).
In C, we store the parity operator
and the eigenvalues of $\mathcal{J}_y$ for convenience (see Sec.~\ref{sourceCode}).
Estimated times for $d=1000$ are $16$ min (C) and $21$ s (D). 
\label{storage}
}

\begin{tabular}{@{\hspace{-1mm}} c @{\hspace{3mm}} l @{\hspace{2mm}} r @{\hspace{6mm}} c @{\hspace{6mm}} c @{\hspace{6mm}} c  @{\hspace{3mm}}}
\hline\hline
\\[-3mm]
& \multicolumn{5}{l}{Method C: matrices $K_\ell$ are computed on the fly}\\ 
& & Dim.   & Disk Storage  & RAM & Time\\[0mm]
& & d & $\mathcal{O}(d^2)$ & $\mathcal{O}(d^2)$ & $\mathcal{O}(d^4)$
\\[0mm]  \hline \\[-1em]
& &
\begin{tabular}{@{} r @{}}
10 \\
50 \\
100 \\
200 \\
500 \\
1000 \\
\end{tabular} 
&
\begin{tabular}{@{} c @{\hspace{1mm}} c @{}}
1.76 & kB\\
40.8 & kB\\
161  & kB\\
643  & kB\\	
4.00 & MB\\
16.0 & MB\\
\end{tabular}
&
\begin{tabular}{@{} c @{\hspace{1mm}} c @{}}
8.98 & kB\\
236 & kB\\
953 & kB\\
3.82 & MB\\
23.9 & MB\\
95.9 & MB\\
\end{tabular}
&
\begin{tabular}{@{} c @{\hspace{1mm}} c @{}}
15.4 & $\mu $s\\
8.18 & ms\\
122 & ms\\
1.81 & s\\
1.07 & min\\
\\
\end{tabular}\\[12mm]
& \multicolumn{5}{l}{Method D: matrices $K_\ell$ have been precomputed}\\ 
& & Dim.   & Disk Storage  & RAM & Time\\[0mm]
& & d & $\mathcal{O}(d^3)$ & $\mathcal{O}(d^2)$ & $\mathcal{O}(d^3)$
\\[0mm]  \hline \\[-1em]
& &
\begin{tabular}{@{} r @{}}
10 \\
50 \\
100 \\
200 \\
500 \\
1000 \\
\end{tabular} 
&
\begin{tabular}{@{} c @{\hspace{1mm}} c @{}}
30.4 & kB\\
3.96 & MB\\
31.8  & MB\\
255  & MB\\
3.99 & GB\\
31.9 & GB\\
\end{tabular}
&
\begin{tabular}{@{} c @{\hspace{1mm}} c @{}}
8.97 & kB\\
236 & kB\\
953 & kB\\
3.82 & MB\\
23.9 & MB\\
95.9 & MB\\
\end{tabular}
&
\begin{tabular}{@{} c @{\hspace{1mm}} c @{}}
2.89 & $\mu $s\\
720 & $\mu $s\\
7.75 & ms\\
83.4 & ms\\
1.93 & s\\
\\
\end{tabular}
\\[1mm]  \hline \hline  \\[-6mm]
\end{tabular}
\end{table}

The second algorithm to compute the Fourier coefficients in Eq.~\eqref{PSfctFourier} is denoted
as Method~D: The matrices $K_\ell$ are precomputed
for every $\ell \in \{-J, \dots J\}$ via Eq.~\eqref{precaleq}
and then stored on disk for later use. 
This requires $\mathcal{O}(d^3)$ disk storage
and $\mathcal{O}(d^4)$ precomputation time.
The stored matrices $K_\ell$ are used to sum
Eq.~\eqref{fouriercoeffsmain} in only $\mathcal{O}(d^3)$ time.
This results in a significantly faster implementation (see Fig.~\ref{timesfigure})
which also suggests a better asymptotic scaling
(with a smaller slope in Fig.~\ref{timesfigure}).
The disk storage and RAM requirements for Methods~C and D
are detailed in Table~\ref{storage} while assuming double precision.
Method~D is preferable (at least) for dimensions $d \leq 500$
as  it significantly reduces the run time with a reasonable amount of disk storage.
For larger dimensions, one has to balance speed with storage requirements.

\subsection{Spherical sampling of the phase-space function via
a fast Fourier transform \label{fftsec}}

We now utilize the Fourier series from
Sec.~\ref{main}
to obtain an equiangular spherical sampling 
of a phase-space function
by applying a fast Fourier transform. We start with the
$(4J{+}1)\times(4J{+}1)$ Fourier coefficients $F_{\ell m}$ from
Eq.~\eqref{PSfctFourier} and Result~\ref{result1} and recall
that the spherical phase-space functions
are band limited with frequency components between $-2J$ and $2J$.
The fast Fourier transform has in this case
an asymptotically negligible $\mathcal{O}(d^2 \log^2(d))$ time complexity 
and results in a grid 
with $(4J{+}1)\times(4J{+}1)$ spherical samples of the phase-space function.
But this is only the coarsest grid possible 
for a complete reconstruction (refer to Eq.~\ref{sphsamples}) 
and finer girds can correct for non-uniformities
and
lead to smoother spherical representations.

In order to obtain a finer grid, it is preferable to add zero padding to the Fourier coefficients
which results in a $n \times 2n$  coefficient array with additional zeros where
$n \geq 4J{+}2$. Many FFT implementations are optimized for $n$ being a power of two.
After applying the FFT, one essentially obtains two copies of the phase-space function
as $\theta$ varies over $0 \leq \theta < 2\pi$ in the result
(while the phase-space function is only defined for $0 \leq \theta < \pi$).
However, by straightforwardly discarding the redundant half one recovers the desired $n \times n$ sampling
of the phase-space function. 

Note that this equiangular sampling is compatible 
with (equiangular) spherical harmonics transforms (see Sec.~\eqref{sphsamples} and, e.g., \cite{driscoll94,kennedy2013book,libsharp}) that could be used to compute the coefficients 
$c_{jm}$ in Eq.~\eqref{sphdecomp}. We also remark that performing fast Fourier transforms
is usually preferable to fast spherical transforms (which are used in Methods~A and B).
This is particularly relevant when one aims at sampling 
phase-space functions for a fixed dimension $d$ to an arbitrarily high resolution $n$.
The two-dimensional FFT takes $\mathcal{O}(n^2 \log^2(n))$ time. Practical
spherical harmonics transforms have, however, a time complexity between 
$\mathcal{O}(n^{5/2} \log(n))$ and $\mathcal{O}(n^{3})$ depending on the implementation
\cite{suda2002fast,driscoll94,libsharp,shtns,mohlenkamp1999fast} and asymptotically faster
implementations might introduce numerical errors and only become superior for 
very fine resolutions \cite{shtns}.

\subsection{Implementations of our algorithms \label{sourceCode}}
We have made implementations of our algorithms for computing spherical samplings
of phase-space functions freely available \cite{gitcode}.
The algorithm for precomputing the coefficients $K_\ell$ 
in Eq.~\eqref{precaleq} for a fixed dimension $d$ has been implemented in C
without any external dependencies. For convenience, we provide a program
(with external dependencies as LAPACK~\cite{lapack})
 to
precompute the parity operators $[M_s]_{\xi \xi}$ and eigenvectors $| U_\nu \rangle $
(Sec.~\ref{eigvecsec}), even though their computation time and storage requirements are negligible (see
Table~\ref{storage}). We currently interface with the precomputed data for $d\leq 500$.
Using the precomputed data,
implementations of Method~D with suitable zero padding 
(Sec.~\ref{fftsec})
are available for C, MATLAB, Mathematica, and Python
\footnote{The current implementation of Method~D has an additional bottleneck
as it reads all of the disk storage
into RAM when computing a phase-space function. For large dimensions as $d \geq 1000$,
this can be avoided without affecting the efficiency of our implementation by
reading the matrices sequentially.}.

\section{Discussion\label{discussion}}

Traditional approaches to efficiently compute spherical
phase-space functions
rely heavily on 
expensive evaluations of Clebsch-Gordan coefficients
and use spherical harmonics transformations
(see Sec.~\ref{prior}). 
We provide much faster algorithms by going beyond
these techniques and by applying a suitable Fourier expansion
and a fast Fourier transform.
This leads to the two variants 
(Method C and D)
which involve different 
time-memory tradeoffs.
Method C calculates the transformation matrices $K_\ell$
on-the-fly and they are then employed to spherically sample
the phase-space function in $\mathcal{O}(d^4)$ time. Method D
precomputes the transformation matrices $K_\ell$ and stores
them using $\mathcal{O}(d^3)$ disk space.
The stored transformation matrices enable us 
to spherically sample the phase-space functions in $\mathcal{O}(d^3)$ time.
We have implemented our algorithms in various 
programming environments
such as C, MATLAB, Mathematica, and Python \cite{gitcode}.

We also remark that our C implementation can be further optimized,
e.g., with regard to memory handling and loops.
The overall run time of the discussed algorithms
could be reduced by truncating spherical-harmonics or Fourier 
coefficients which could be motivated by prior knowledge
or symmetry considerations.
In addition, the disk storage of Method D can be optimized
to $\mathcal{O}(d)$ if the summation in Eq.~\eqref{PSfctFourier} 
can be restricted to Fourier coefficients $F_{\ell m}$ 
with $\ell,m \leq t$ for some suitable constant $t$.
But this might not be a good approximation for general quantum states
and we are focussing on computing phase-space function
exactly up to numerical precision.

We finally discuss how our results could
be applied to compute analytical derivatives
with respect to spherical rotation angles. Following Sec.~\ref{mainsec}
and Result~\ref{result1},
one obtains the Fourier coefficients $F_{\ell m}$ and this 
representation helps us to
compute derivatives analytically
by multiplying the coefficients $F_{\ell m}$ with $i \times \ell$ (or 
$i \times m$): 
\begin{align*}
\partial_\theta F_\rho (\theta,\phi,s) &= \sum_{\ell,m= -2J}^{2J} 
e^{i \ell \theta}  e^{i m \phi} \,  i \, \ell \, F_{\ell m},\\
\partial_\phi F_\rho (\theta,\phi,s) &= \sum_{\ell,m= -2J}^{2J} 
e^{i \ell \theta}  e^{i m \phi} \,  i \, m \, F_{\ell m}.
\end{align*}
These derivatives are particularly relevant for the computation of
star products of phase-space functions (see \cite{KZG}).
This can be extended to analytical gradients
\begin{equation*}
\mathrm{grad}[F_\rho (\theta,\phi,s)] = (
\partial_\theta F_\rho (\theta,\phi,s),
\partial_\phi F_\rho (\theta,\phi,s)
),
\end{equation*}
which enables us to search for local extrema of phase-space functions
(e.g., minima of locally negative regions)
via gradient descent optimizations.

\section{Conclusion\label{conclusion}}
In this work, we have considered spherical phase spaces 
of large quantum states
and have provided effective computational methods for them.
Our methods allow now for 
much larger dimensions than before. Going beyond
approaches using
tensor-operator decompositions and spherical-harmonics transforms,
we can directly harness the efficiency of the fast Fourier transform
applied to an efficiently computable Fourier series expansion.
Our C implementation \cite{gitcode} is at least an order of magnitude 
faster than prior implementations when compared for up to
dimension 500 (or up to 499 qubits in permutationally
symmetric states). Our data  also suggest an asymptotic speed-up
by utilizing suitable precomputations.

The presented computational methods for phase spaces
of single-qudit and permutation-symmetric multi-qubit states
have applications to many-body physics, quantum metrology, and 
entanglement validation. We have illustrated many-body
examples in Sec.~\ref{applications} some of which are pursued 
in current quantum hardware. Our results will enable
both theoreticians and experimentalists to more effectively work with
phase-space representations in order to study high-dimensional quantum effects.
This will help to 
guide future experimental advancements in generating
complex quantum states of high fidelities
\cite{google_supr, catstate, Song574, preskill2018quantum}.

\begin{acknowledgments}
	B.~Koczor acknowledges financial support from the European 
	Union’s Horizon 2020 research and innovation programme 
	under Grant Agreement No.~820495 (AQTION).
	This work is supported in part by the Elitenetzwerk Bayern through ExQM
	and the Deutsche Forschungsgemeinschaft (DFG, German Research Foundation) 
	under Germany’s Excellence Strategy -- EXC-2111 -- 39081486.
	R.~Zeier acknowledges funding from funding from the European 
	Union’s Horizon 2020 research and innovation programme under 
	Grant Agreement No.~817482 (PASQuanS).
\end{acknowledgments}

\appendix

\section{Computing tensor-operator decompositions \label{topdecomp}}

One can obtain phase-space functions via the tensor-operator decomposition in Eq.~\eqref{sphdecomp}.
This requires the evaluation of $\mathcal{O}(d^2)$ operations as $c_{jm} = \Tr \,[ \rho\,  \T_{jm}^\dagger ]$.
Tensor operators can be specified in terms of Clebsch-Gordan coefficients via Eq.~\eqref{topelements},
but most of their matrix elements are zero due to condition the $C^{J m_1}_{J m_2, j m} = 0$ for $m_1-m_2 \neq m$.
Even though a tensor operator is sparse in this representation due to its $\mathcal{O}(d)$
non-zero elements, obtaining all decomposition coefficients $c_{jm}$ still requires the numerical
evaluation of overall $\mathcal{O}(d^3)$ Clebsch-Gordan coefficients.
This can be seen by expressing the trace explicitly as
\begin{equation*}
c_{jm} = \Tr \,[ \rho\,  \T_{jm}^\dagger ] = \sum_{m_1 = -J}^J  [\rho]_{m_1, m_1+m} [\T_{jm}]_{m_1+m, m_1}  
\end{equation*}
where we have used the condition $[\T_{jm}]_{m_1 m_2} = 0$
if $m_1-m_2 \neq m$. It is clear from the above summation that computing all the coefficients $c_{jm}$
requires one to evaluate  $\mathcal{O}(d^3)$ Clebsch-Gordan coefficients as the matrix
elements $[\T_{jm}]_{m_1+m, m_1}$. The elements $[\rho]_{m_1, m_1+m}$ should be directly
available in memory and the overall computation time of this approach is therefore dominated by evaluating the
Clebsch-Gordan coefficients. We expect that computing a single one of them requires
$\mathcal{O}(d^n)$ time with $ n > 0$ and based on our numerical computations in Fig.~\ref{timesfigure}
we speculate that $n\approx1$.

\section{Fourier series representation of the rotation operator\label{appendixRotationFourier}}

We now establish how the rotation operator in Eq.~\eqref{PSrepDefinition} can be decomposed
into a Fourier series. This step is crucial for deriving our Result~\ref{result1}, which
finally allows us to efficiently decompose a phase-space function into Fourier components.

Recall that the rotation operator defined in Eq.~\eqref{PSrepDefinition} is parametrized
in terms of Euler angles as $\mathcal{R}(\theta,\phi) = e^{i\phi \mathcal{J}_z} e^{i\theta \mathcal{J}_y}$
via the spin operators $\mathcal{J}_y$ and $\mathcal{J}_z$.
These spin operators are defined via their commutation relations
$[\mathcal{J}_j,\mathcal{J}_k]=i \sum_\ell \epsilon_{jk\ell} \mathcal{J}_\ell$
for $j,k,\ell = x,y,z$ and $\epsilon_{jk\ell}$ is the Levi-Civita symbol, refer to, e.g.,
\cite{messiah1962a,sakurai1995modern}.
For an $N$-qubit system these are proportional to sums of Pauli operators
$\mathcal{J}_j = \tfrac{1}{2} \sum_{k=1}^N \sigma_j^{(k)} $
acting on individual qubits and $j \in \{x,y,z\}$.
These operators are unitarily equivalent and have the
eigenvalues $m \in \{-J, -J+1, \dots, J \}$
due to the eigenvalue equation
\begin{equation} \label{singlespineigenvalue}
\mathcal{J}_y | U_m \rangle := m | U_m \rangle
\quad  \quad
\mathcal{J}_z | J m \rangle := m | J m \rangle.
\end{equation}
Note that in an N qubit system $2J = N$.
Here we denote eigenvectors of the $\mathcal{J}_y $ operator as
$| U_m \rangle$ and recall the orthogonality condition
$\langle  U_m | U_n \rangle = \langle J m | Jn \rangle = \delta_{m n}$.
The spectral resolution of these spin operators is obtained in terms of the 
rank-1 projectors $| U_m \rangle \langle  U_m| =: A_m$ and
$ | J m \rangle\langle  J m| =: B_m$ as
\begin{equation} \label{ioperatordecomps}
\mathcal{J}_y = \sum_{m=-J}^{J} m \, A_m
\quad \quad
\mathcal{J}_z = \sum_{m=-J}^{J} m \, B_m.
\end{equation}
It immediately follows that rotation operators decompose into
the following sum of rank-one projectors
\begin{equation} \label{spinopdecomp}
e^{i\theta \mathcal{J}_y}= \sum_{m=-J}^{J} e^{i \theta m} \, A_m
\quad \quad
e^{i\phi \mathcal{J}_z}  = \sum_{m=-J}^{J} e^{i \phi m} \, B_m.
\end{equation}
Note that the dependency on the rotation angles $\theta$ and $\phi$
is now completely absorbed by the Fourier components $e^{i \theta m}$ and $e^{i \phi m}$.

The rank-1 matrices $A_m$ and $B_m$ are projections onto the eigenvectors
of the spin operator $\mathcal{J}_y$ from Eq.~\eqref{ioperatordecomps}
and we define their matrix elements as
\begin{eqnarray}
[A_m]_{m_1 m_2} = \langle J m_1 | A_m | J m_2\rangle,
\end{eqnarray}
and trivially $[B_m]_{m_1 m_2} = \delta_{m_1 m_2}$.

Matrix elements of $A_m$ have been used in \cite{tajima2015analytical,wigdmatrix2}
for efficiently computing 
Wigner-d matrices via the Fourier series decomposition
\begin{align} \label{fourierrep} 
d_{m_1,m_2}^J(\theta):=& \nonumber
\langle J m_1| e^{i\theta \mathcal{J}_y} | J m_2\rangle\\
=& \sum_{m=-J}^{J} e^{i \theta m} \, [A_m]_{m_1 m_2}.
\end{align}
Note that here $[A_m]_{m_1 m_2}$ appear as Fourier series decomposition
coefficients of the Wigner-d matrix elements. This form was originally
proposed in \cite{tajima2015analytical} for 
efficiently calculating $d_{m_1,m_2}^J(\theta)$ via fast Fourier transforms
as the advantage of this representation is that the summation in Eq.~\eqref{fourierrep}
is numerically stable due to the boundedness of the matrix elements as
$|[A_m]_{m_1 m_2}| \leq 1$. Instead of computing Wigner-d matrix elements,
our approach in Result~\ref{result1} relies directly on the matrices  $A_m$.

\subsection{Analytical expression for  $[A_m]_{m_1 m_2}$}
The explicit form of the Fourier coefficients  $[A_m]_{m_1 m_2}$ was derived 
analytically in \cite{tajima2015analytical}  as
\begin{equation*} 
[A_m]_{m_1 m_2} = \sum_{k=a}^{b}
w_k^{(m_1 m_2)} \, I_m(J,2k+m_1-m_2)
\end{equation*}
with summation bounds $a=\max{(0,m_2-m_1)}$ and $b=\min{(J-m_1,J+m_2)}$.
The explicit form of the coefficients appearing in the
above summation are 
\begin{align*}
w_k^{(m_1 m_2)} =& (-1)^{k+m_1-m_2}\\ & \times \frac
{
	\sqrt{(J+m_1)! (J-m_1)! (J+m_2)! (J-m_2)!}
}
{(J-m_1-k)! (J+m_2-k)! (k+m_1-m_2)! k!},\\
I_m(J,\lambda) =& 2^{-2J}\sum_{\ell=c}^d (-1)^{\ell-\lambda/2}
\binom{2J-\lambda}{J+m-\ell} \binom{\lambda}{\ell}
\end{align*}
with summation bounds $c=\max{(0,-J+m+\lambda)}$
and here $d=\min{(\lambda,J+m)}$ and $(\dots)!$ denotes
the factorial function while $\binom{\lambda}{\ell}$ are
the binomial coefficients.

\subsection{Numerical computation of the eigenvectors} \label{eigvecsec}
A simple and efficient way for numerically evaluating the coefficients
$[A_m]_{m_1 m_2}$ in Eq.~\eqref{fourierrep} was proposed in \cite{wigdmatrix2}.
This approach first computes the eigenvectors $| U_m \rangle$ from Eq.~\eqref{singlespineigenvalue}
by numerically diagonalizing the spin operator $\mathcal{J}_y$.
One then obtains the numerical representation of the eigenvectors
$| U_m \rangle $ that define the rank-1 projector $A_m = | U_m \rangle \langle  U_m|$.
Its matrix elements can then be obtained straightforwardly
\begin{equation} \label{eigenvectors}
[A_m]_{m_1 m_2} = [U_m]_{m_1} [U_m]_{m_2}^*
\end{equation}
as products of vector entries of 
eigenvectors of
$\mathcal{J}_y$
from Eq.~\eqref{singlespineigenvalue}
and here $[\dots]^*$ denotes complex conjugation.
The matrix $\mathcal{J}_y$ can be diagonalised to numerical precision
(it is tridiagonal and Hermitian) which provides a
high-precision numerical representations of $[A_m]_{m_1 m_2}$.
This has been demonstrated in \cite{wigdmatrix2}
using the ZHBEV diagonalisation routine of the software
package LAPACK \cite{lapack}. We use this approach in this work for numerically
computing eigenvectors.

\section{Derivation of Result~\ref{result1}} \label{derivation}
Substituting the expansion of rotation operators from Eq.~\eqref{spinopdecomp}
into our definition of phase spaces in Eq.~\eqref{PSrepDefinition} and using that the rank-one
projectors $A_m$ and $B_m$ are self adjoint we obtain
\begin{align}
&F_\rho (\theta,\phi,s) =  \Tr \,[ \, \rho \, e^{i\phi \mathcal{J}_z} e^{i\theta \mathcal{J}_y}  M_s  e^{-i\theta \mathcal{J}_y} e^{-i\phi \mathcal{J}_z}  ]\\
&= \sum_{\mu, \nu, \kappa, \lambda = -J}^J \label{fdecomp}
e^{i (\kappa-\lambda) \phi} e^{i (\mu-\nu) \theta} \Tr \,[ \, \rho \, B_\kappa A_\mu  M_s A_\nu B_\lambda ].
\end{align}
This is a Fourier series decomposition of the phase-space functions.
It is our aim now to express its Fourier
coefficients explicitly.
In particular, one can rearrange the terms in the trace and obtain
\begin{equation*}
\Tr \,[ \, \rho \, B_\kappa A_\mu  M_s A_\nu B_\lambda ] = \Tr \,[ \,  B_\lambda\, \rho \, B_\kappa A_\mu  M_s A_\nu],
\end{equation*}
where the first term in the trace is simply a projection of the density matrix onto a single matrix element
in the $z$ basis as $ B_\lambda\, \rho \, B_\kappa
= | J \lambda \rangle \langle  J \kappa| \rho_{\lambda\kappa}$. Here, matrix elements of the density
operator are denoted as $ \rho_{\lambda\kappa}:= \langle J \lambda | \rho | J \kappa \rangle$ assuming the standard $z$ basis.
Now the Fourier components
$\Tr \,[ \, \rho \, B_\kappa A_\mu  M_s A_\nu B_\lambda ]=\rho_{\lambda\kappa}  \Tr \,[ \, | J \lambda \rangle \langle  J \kappa| A_\mu  M_s A_\nu]$
in Eq.~\eqref{fdecomp} can be simplified into the form $\rho_{\lambda\kappa}  \, \langle  J \kappa| A_\mu  M_s A_\nu  | J \lambda \rangle$
which is a product of single matrix elements in the standard $z$ basis as
\begin{equation*}
\Tr \,[ \, \rho \, B_\kappa A_\mu  M_s A_\nu B_\lambda ]
=\rho_{\lambda\kappa}  \,[A_\mu  M_s A_\nu]_{\kappa\lambda}.
\end{equation*}
Equation~\eqref{fdecomp} finally reads
\begin{equation*}
F_\rho (\theta,\phi,s) = \sum_{\mu, \nu, \kappa, \lambda = -J}^J 
e^{i (\kappa-\lambda) \phi} e^{i (\mu-\nu) \theta} \, \rho_{\lambda\kappa}  \, [A_\mu  M_s A_\nu]_{\kappa\lambda}.
\end{equation*}
We now explicitly express this phase-space function as a Fourier series and denote its
expansion coefficients as $F_{\ell m}$ via
\begin{equation*}
F_\rho (\theta,\phi,s) = \sum_{\ell,m= -2J}^{2J} 
e^{i m \phi} e^{i \ell \theta} F_{\ell m}.
\end{equation*}
The expansion coeffiecents are given by a finite sum 
using the new indexes $\mu \rightarrow \nu +\ell$ and $\kappa \rightarrow \lambda+m$, it follows
\begin{equation*}
 F_{\ell m} = \sum_{\substack{\nu, \lambda = -J \\ -J \leq (\nu +\ell),(\lambda+m) \leq J}}^J 
 \rho_{\lambda,\lambda+m}  \, [A_{\nu +\ell}  M_s A_\nu]_{\lambda+m,\lambda}.
\end{equation*}
We slightly simplify the previous equation by applying the transpose of the matrix product
$[A_{\nu +\ell}  M_s A_\nu]_{\lambda+m,\lambda} = [A_{\nu }  M_s A_{\nu + \ell}]_{\lambda,\lambda+m}$,
which results in our final expression
\begin{equation*}
F_{\ell m} = \sum_{\substack{\lambda = -J \\ -J \leq (\lambda+m) \leq J}}^J 
\rho_{\lambda,\lambda+m}  \, [K_\ell]_{\lambda,\lambda+m}.
\end{equation*}
Here we have introduced the set of matrices $K_\ell$ which simply
multiply the density matrix element-wise
and we define their explicit form as
a summation over the matrix products
\begin{equation} \label{coefficeintdef}
K_\ell := \sum_{\substack{\nu = -J \\ -J \leq (\nu +\ell)\leq J}}^J    \, A_{\nu }  M_s A_{\nu + \ell}.
\end{equation}
Note that the Fourier coefficients $F_{\ell m}$ depend both on the density operator $\rho$
and on the parity operator $M_s$, and implicitly on the eigenvectors of $\mathcal{J}_y$.
We have  introduced the matrices $K_\ell$,
which completely determine the dependence on the parity operator and on the eigenvectors of
$\mathcal{J}_y$. These matrices can be precomputed and stored or computed on-the-fly. 
The Fourier coefficients can then be completely determined via the
efficient summation
\begin{equation} \label{mainresultappendix}
F_{\ell m} = \sum_{\substack{\lambda = -J \\ -J \leq (\lambda+m) \leq J}}^J 
[\rho \circ K_\ell]_{\lambda,\lambda+m}
\end{equation}
of the element-wise matrix products $[\rho \circ K_\ell]$.

\section{Calculating the transformation matrices $K_\lambda$} \label{precalc}

The coefficient matrices in Eq.~\eqref{coefficeintdef} can be calculated
efficiently by using the earlier definition
$| U_m \rangle \langle  U_m| =: A_m$, which results in  
\begin{equation*} 
K_\ell = \sum_{\substack{\nu = -J \\ -J \leq (\nu +\ell)\leq J}}^J    \, | U_\nu \rangle \langle  U_\nu|  M_s | U_{\nu + \ell} \rangle \langle  U_{\nu + \ell}|.
\end{equation*}
We define the basis-transformed parity operator
$\tilde{M}_s := U M_s U^{\dagger} $ using the unitary
operator $U$ whose column vectors are composed of the eigenvectors $| U_\nu \rangle$
-- and which diagonalizes $\mathcal{J}_y$ as discussed in Appendix~\ref{appendixRotationFourier}.
The expression for computing the matrices simplifies to the form
\begin{equation} \label{precalceq}
K_\ell = \sum_{\substack{\nu = -J \\ -J \leq (\nu +\ell)\leq J}}^J  [\tilde{M}_s ]_{\nu,\nu + \ell}  \, | U_\nu \rangle \langle  U_{\nu + \ell}|.
\end{equation}
We evaluate this expression numerically by first computing eigenvalues and eigenvectors of the $y$
component of the angular momentum operator as discussed in Sec.~\ref{eigvecsec}. This step requires $\mathcal{O}(d^3)$ time
where $d=2J+1$.  We than compute and basis transform the parity operator to obtain $\tilde{M}_s$, which requires
$\mathcal{O}(d^3)$ time (via a naive matrix multiplication algorithm)
and storing the result requires $\mathcal{O}(d^2)$ space.

We now fix $\ell$ and evaluate Eq.~\eqref{precalceq} for this fixed $\ell$.
We compute the matrix $K_\ell$ element-wise as $[K_\ell]_{ab}$ using the explicit expression
 $[| U_\nu \rangle \langle  U_{\nu + \ell}|]_{ab} =  [U]_{\nu a} ([U]_{\nu+\ell,b}])^*$,
where $^*$ denotes complex conjugation.
Computing such a matrix $K_\ell$ in Eq.~\eqref{precalceq} requires $\mathcal{O}(d^3)$ time for a fixed $\ell$.
We therefore conclude that computing every coefficient matrix $K_\ell$ with $\ell \in \{-2J,  \dots 2J\}$
requires $\mathcal{O}(d^4)$ time.

After computing $K_\ell$ for a fixed $\ell$, one can proceed according to two distinct strategies, which we refer
to as Method C and D in the main text.
In case of Method D, we store the matrix $K_\ell$ and repeat this procedure for each $\ell \in \{-2J,  \dots 2J\}$.
This requires $\mathcal{O}(d^3)$ disk storage space. These precomputed matrices can be used later 
in Result~\ref{result1} for computing phase spaces in $\mathcal{O}(d^3)$ time which requires only
$\mathcal{O}(d^2)$ memory, i.e., for $\rho$,  $U$ and $\tilde{M}_s$, and one only reads in a single matrix $K_\ell$ at a time.
In case of Method C, we compute $K_\ell$ for a fixed $\ell$, and use it immediately for 
evaluating the summation in Result~\ref{result1} for a fixed $\ell$. We can then repeat this procedure for
each $\ell \in \{-2J,  \dots 2J\}$. Therefore, Method C does not require disk storage space for the matrices $K_\ell$,
but allows for calculating phase-spaces via Result~\ref{result1} in $\mathcal{O}(d^4)$ time and
similarly using $\mathcal{O}(d^2)$ memory.

%\bibliography{references}


\end{document}